\documentclass[aps,prb,twocolumn,superscriptaddress]{revtex4-1}

\usepackage{mathrsfs}
\usepackage{amsmath}
\usepackage{amssymb}
\usepackage{amstext}
\usepackage{graphicx, epsf}
\usepackage{enumerate}
\usepackage{bbm}
\usepackage{color}
\usepackage[english]{babel}

\begin{document}

\title{Scattering strength of potassium on a carbon nanotube with known chirality}

\author{Ryuichi Tsuchikawa}

\affiliation{Department of Physics, University of Central Florida,
  Orlando, Florida 32816, USA}

\affiliation{Nanoscience Technology Center, University of Central
  Florida, Orlando, FL 32816, USA}

\author{D. Heligman}

\affiliation{Department of Physics, University of Central Florida,
  Orlando, Florida 32816, USA}

\affiliation{Nanoscience Technology Center, University of Central
  Florida, Orlando, FL 32816, USA}

\author{Z. Y. Zhang}

\affiliation{Department of Mechanical Engineering, Columbia
  University, New York, NY 10027, USA}

\author{A. Ahmadi}

\affiliation{Department of Physics, University of Central Florida,
  Orlando, Florida 32816, USA}

\author{E. R. Mucciolo}

\affiliation{Department of Physics, University of Central Florida,
  Orlando, Florida 32816, USA}

\author{J. Hone}

\affiliation{Department of Mechanical Engineering, Columbia
  University, New York, NY 10027, USA}

\author{M. Ishigami}

\affiliation{Department of Physics, University of Central Florida,
  Orlando, Florida 32816, USA}

\affiliation{Nanoscience Technology Center, University of Central
  Florida, Orlando, FL 32816, USA}

\date{\today}


\begin{abstract}
We have measured the scattering strength of charged impurities on a semiconducting single-walled carbon nanotube with known chirality. The resistivity of the nanotube is measured as a function of the density of adsorbed potassium atoms, enabling the determination of the resistance added by an individual potassium atom. Holes are scattered 37 times more efficiently than electrons by an adsorbed potassium atom. The determined scattering strength is used to reveal the spatial extent and depth of the scattering potential for potassium, a model Coulomb adsorbate. Our result represents an essential experimental input to understand adsorbate-induced scattering and provides a crucial step for paving the way to rational design of nanotube-based sensors.
\end{abstract}

\maketitle


The well-established utility of carbon nanotubes in sensing applications \cite{ref1,ref2} arises from their high sensitivity to adsorbed species. The impact of adsorbates on the transport properties of nanotubes has been extensively studied theoretically \cite{add1, add2, ref3, ref4, ref5, ref6, ref7, ref8, ref9, ref10} and exceptional sensitivities down to single adsorbates for certain species have been predicted. Such sensitivities are expected because nanotubes are one-dimensional (1D) conduction channels where any uncorrelated disorder induced by adsorbates can induce drastic effects, including charge localization. Previous theoretical studies \cite{add1, add2, ref3, ref4, ref5, ref6, ref7, ref8, ref9, ref10} have identified the dependence of the resistance induced by external scattering potentials on the chirality of nanotubes as well as on the depths and spatial extents of the potentials. However, experiments have not been performed to verify these calculations because of the difficulty in carrying out well-controlled measurements on nanotubes with known chirality. Furthermore, it is very challenging to precisely calculate the nature of scattering potentials induced by adsorbates because of the complex nature of screening in nanotubes. Indeed, previous calculations of screening effects \cite{ref11, ref12, ref13} have focused on their impact on electron-electron interactions rather than electron-adsorbate interactions and are not necessarily applicable to the realistic modeling of adsorbates. As a result, model scattering potentials used in the previous calculations have no connections to the actual effective potential exerted by adsorbates and the impact of adsorbates remains unknown. Therefore, measurements of resistance induced by adsorbates on nanotubes with known chirality are still needed to establish the fundamental science of nanotube-based sensors and to determine the ultimate potential of nanotubes for sensor technologies.

Most adsorbates are expected to transfer charge to nanotubes and exert a Coulomb-like potential on charge carriers. Previous experiments on the impact of charged adsorbates on individual nanotubes have focused on reducing the Schottky barrier at the contacts \cite{ref14,ref15} or on shifting the Fermi level \cite{ref16, ref17, ref18, ref19,ref20,ref21} and not on the resistance induced by the adsorbates. Furthermore so far, no experiments on the impact of adsorbates on nanotubes with known chirality have been performed. In this paper, we determine the resistance added by a model Coulomb adsorbate, potassium, on a semiconducting single-walled nanotube of known chirality by measuring the resistivity added by adsorbates as a function of coverage in the diffusive transport regime. Precise knowledge of the atomic structure of the nanotube is exploited to estimate the depth and spatial extent of the effective potential exerted by the model Coulomb scatterer via direct comparison of the experimental data with theoretical calculations. As such, this work represents an important scientific step toward the rational engineering of sensors based on nanotubes.

Our experimental and theoretical analyses focus on a (7,6) semiconducting nanotube. Multiple devices are prepared on the nanotube following a previously established procedure \cite{ref22}. The nanotube is grown by using chemical vapor deposition across a 60~$\mu$m slit etched through a silicon wafer. The nanotube chiral indices are determined using Rayleigh scattering spectroscopy \cite{ref23}. The nanotube is then transferred onto a doped Si substrate with 280~nm of thermal SiO$_2$ layer. Multiple contacts are fabricated on the nanotube with channel length ranging from 0.5~$\mu$m up to 8~$\mu$m. Gold is used to make an electrical contact, and all devices are annealed in flowing Ar/H$_2$ gases at 360~$^{\rm o}$C for 3 hours to remove polymer residues from the device fabrication process \cite{ref24} prior to measurements.

All measurements are performed in ultra high vacuum (UHV) to eliminate contributions from other adsorbates. Potassium is deposited by activating an alkali metal dispenser (SAES getters), and the incoming flux is controlled with a mechanical shutter. The temperature of the nanotube is maintained below 20~K during the transport measurements in order to prevent surface diffusion of potassium \cite{ref25}. The deposition rate of potassium is measured with a retractable quartz crystal microbalance thickness monitor positioned between the potassium dispenser and the nanotube immediately before deposition. The geometric factor, required to calculate the density of potassium at the nanotube, is separately measured using the same thickness monitor. The sticking coefficients of potassium on the QCM as well as the nanotube at low temperatures are assumed to be unity following previous surface science studies \cite{ref26} of adsorption of alkali metals on graphite. Finally, the density of potassium on the nanotube is calculated by considering effective area to be diameter $\times$ length. Potassium should become positively charged on the nanotube, introducing both electron doping and additional scattering due to the local potential variation. The added resistance should vary strongly with carrier type: positively charged potassium should create a strongly scattering potential barrier for holes and a more weakly scattering potential well for electrons.


\begin{figure}[t]
\centering
\includegraphics[width=4.25cm]{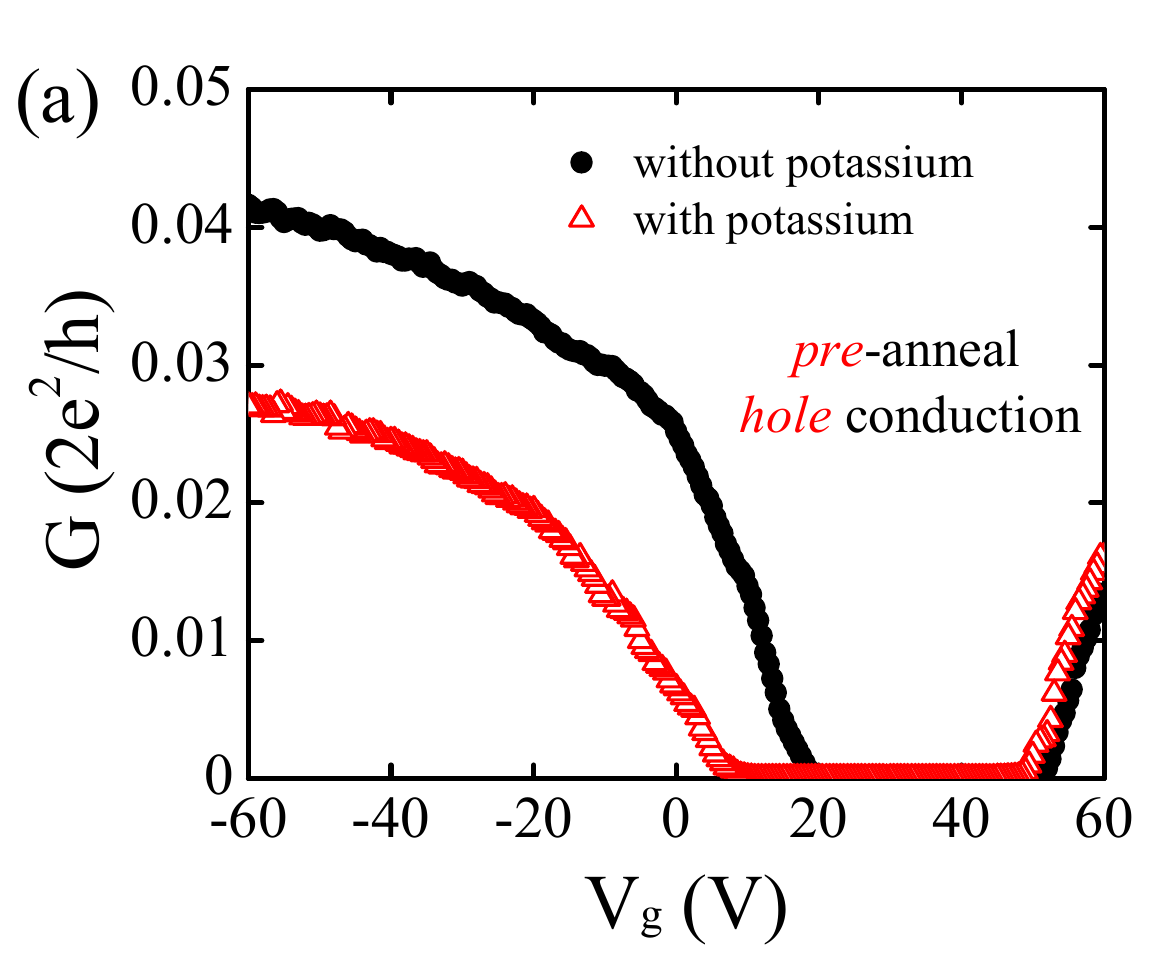}
\includegraphics[width=4.25cm]{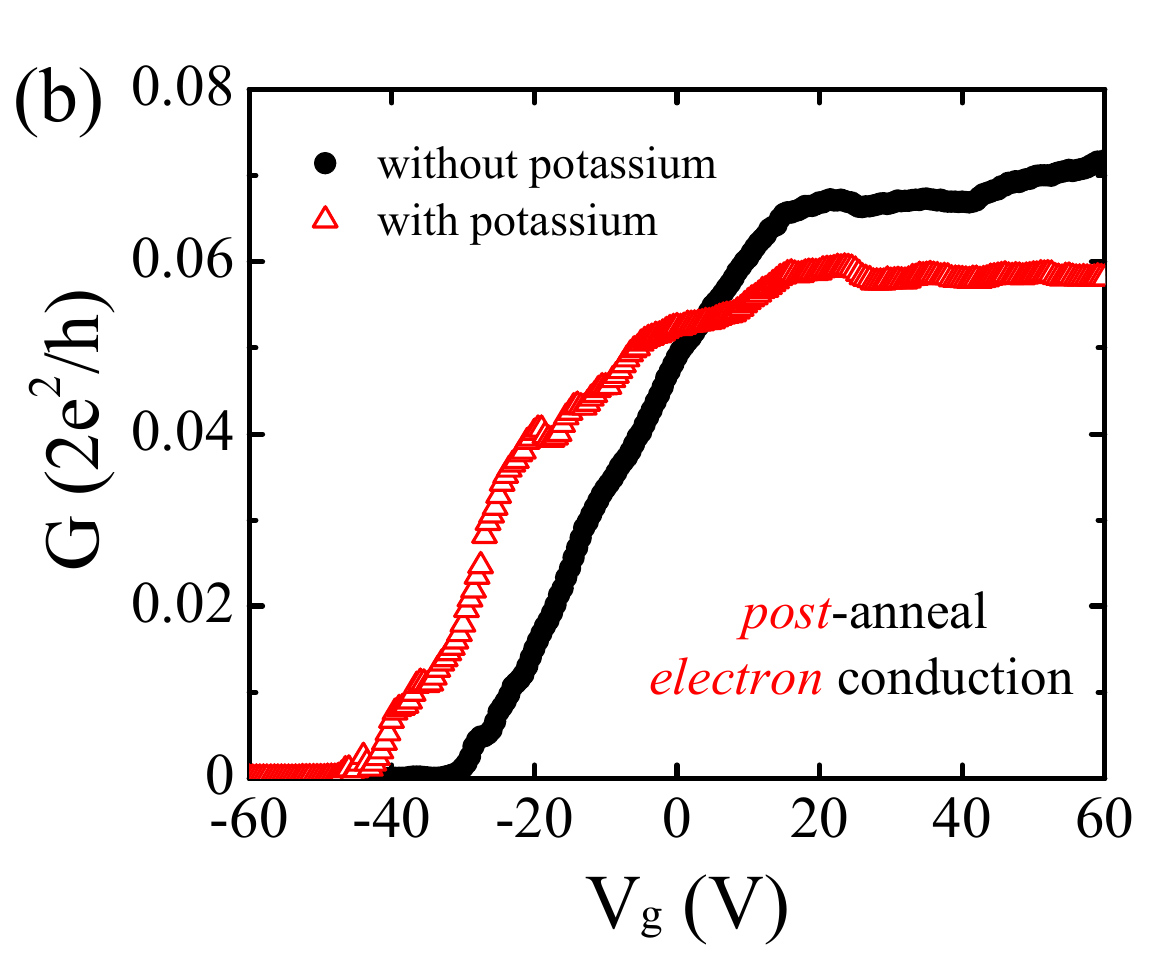}
\caption{(Color online) Conductance as a function of gate voltage for
  a 6~$\mu$m (7,6) nanotube segment in UHV. (a) Hole conductance at
  9~K (pre annealing), and (b) electron conductance at 16~K (post
  annealing) before (black circles) and after (red triangles) dosing
  with potassium.}
\label{fig:1}
\end{figure}


Figure~1 shows the impact of potassium on the two-terminal conductance of a 6~$\mu$m long section of a (7,6) semiconducting nanotube as a function of gate voltage. Upon introduction into the UHV environment, the device shows p-type ambipolar behavior, with hole conduction up to $V_g = 20$~V and electron conduction above $V_g = 50$~V. All devices show similar behavior. As shown in Figure~1a, deposition of $2.2 \pm 0.1$ potassium atoms/$\mu$m suppresses hole conduction, consistent with the notion that adsorbates add scattering. However, it enhances electron conduction, indicating that the device is contact-dominated in this regime and the main effect of the potassium is to modulate the Schottky barrier, as reported previously \cite{ref14, ref21}. The sample is then annealed at 460 K to remove the potassium \cite{ref17}. In order to measure electron rather than hole transport, we anneal the nanotube for several days. As shown in Figure~1b, the device shows n-type ambipolar behavior after this long annealing process. Previous experiment on annealing nanotubes in vacuum \cite{ref21} also resulted in a similar behavior. Such annealing temperature is not expected to induce damages to the nanotube as suggested by a previous Raman spectroscopy study on annealed graphene \cite{ref27}, which showed that annealing in vacuum up to 673 K does not cause appreciable damage to its graphitic lattice. As shown in Figure 1b, deposition of $29.6 \pm 0.4$ potassium atoms/$\mu$m now suppresses electron conduction rather than enhancing it, showing that the channel resistance is now dominating over the contact resistance. These measurements confirm earlier reports that both the contact Schottky barriers and the channel resistance are affected by charged adsorbates such as potassium \cite{ref14, ref18, ref21, ref28, ref29} , and show that two-probe conductance measurements are insufficient for quantifying the impact of adsorbates.

\begin{figure}[t]
\centering
\includegraphics[width=4.25cm]{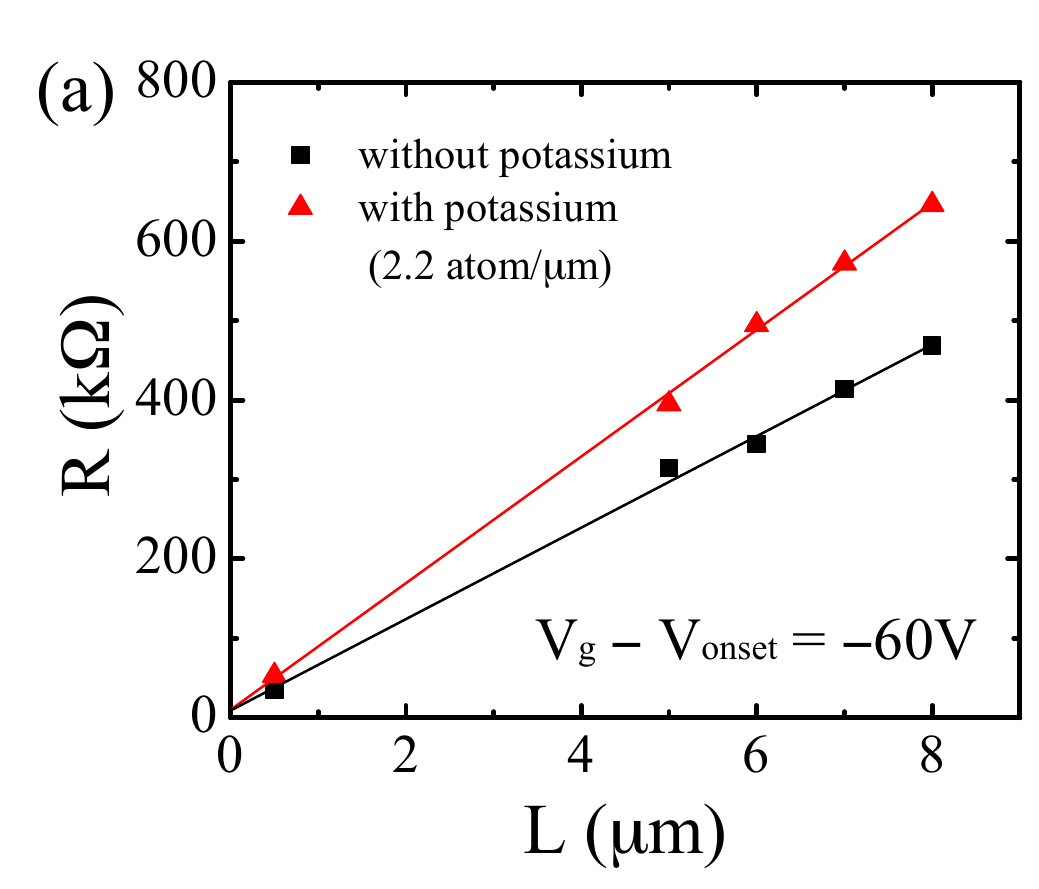}
\includegraphics[width=4.25cm]{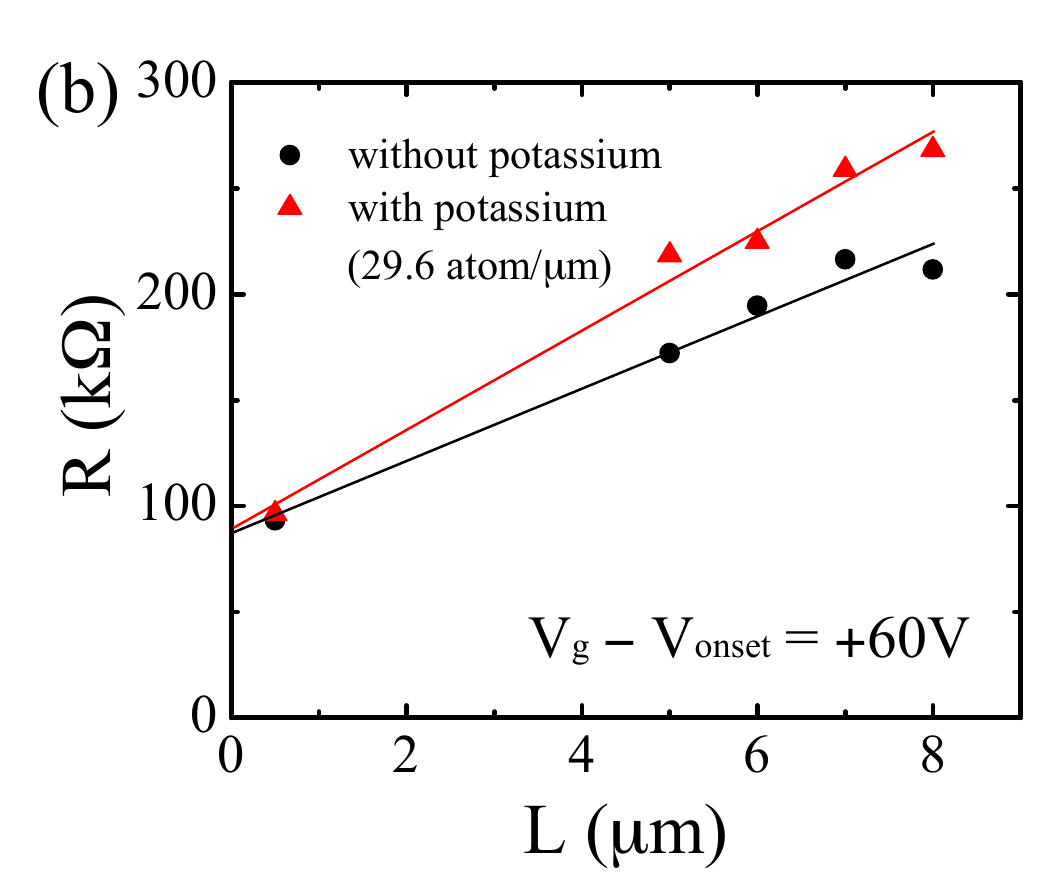}
\includegraphics[width=4.25cm]{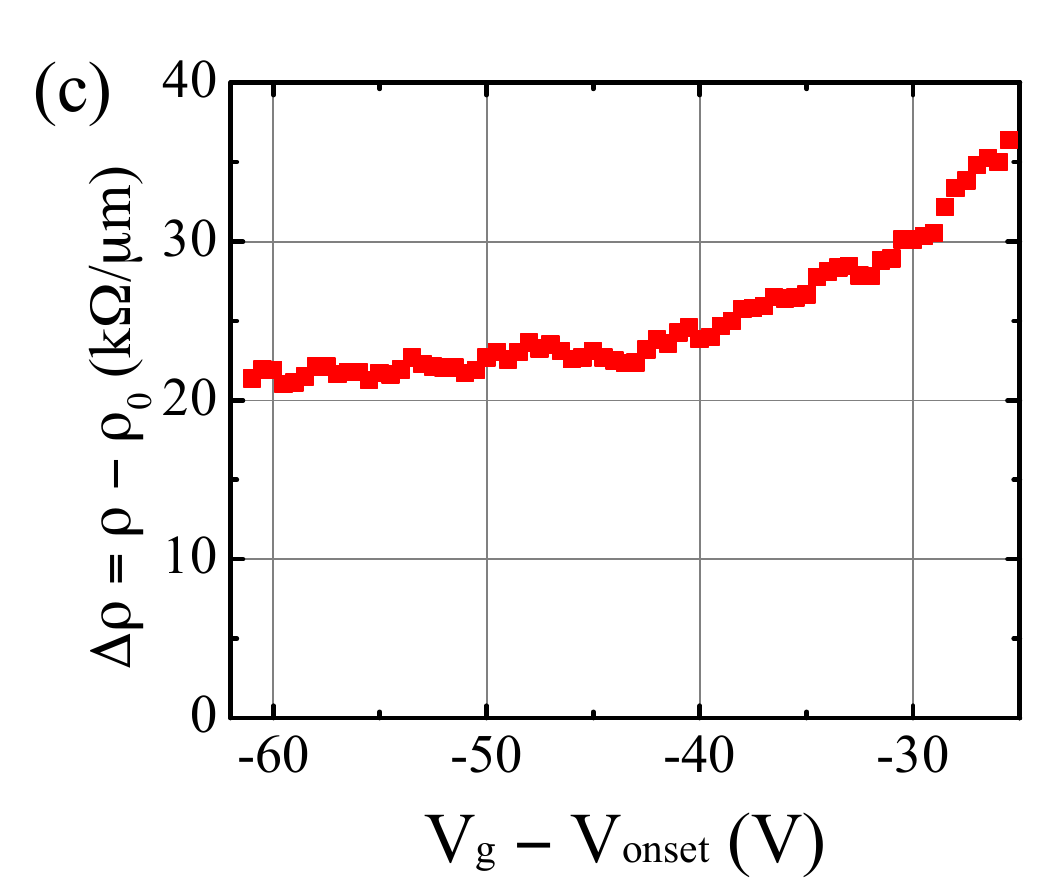}
\includegraphics[width=4.25cm]{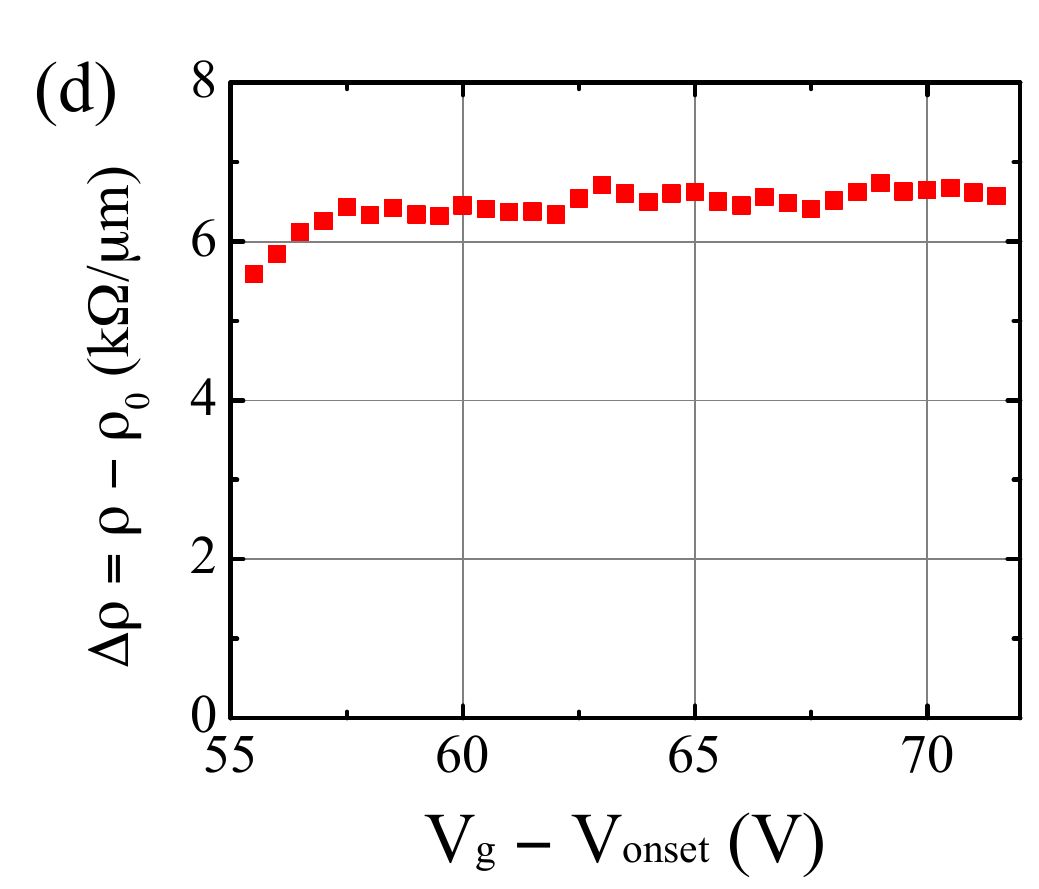}
\caption{(Color online) The length dependence of resistance for
  multiple devices (a) at $V_g - V_{\rm onset} = -60$~V and (b) at
  $V_g - V_{\rm onset} = 60$~V before (black) and after (red) dosing
  with potassium. Gate dependence of resistivity added by potassium
  for (c) holes and (d) electrons.}
\label{fig:2}
\end{figure}

The impact of potassium on the resistivity can be determined by performing length-dependent resistance measurements \cite{ref30, ref31}. Figure 2a shows the resistance as a function of the length in the on-state, $\pm 60$ volts away from the onsets of electron or hole conduction, corresponding to approximately $\pm 1.5$~eV away from the charge neutrality point \cite{ref32}. The solid lines are linear fits, with the contact resistance given by the intercept and the resistivity given by the slope. The resistivity before addition of potassium is $\rho_0 = 57.7 \pm 2.0 $~k$\Omega$/$\mu$m for holes and $\rho_0 = 17.0 \pm 1.6 $~k$\Omega$/$\mu$m  for electrons. The errors in the values of resistivities originate from the linear fits. As such, the errors and error bars discussed and shown below most likely originate from the small, but finite, non-uniformity in contact resistances of different nanotube segments. The larger resistivity for holes indicates that there are more positively charged defect sites on the SiO$_2$ substrate. Such charge traps on SiO$_2$ have been observed in the studies of graphene field effect transistors \cite{ref33,ref34} and our oxygen plasma treatment of the substrate prior to the nanotube transfer process may have imparted more positive charges. While we do not control the initial background impurity level prior to the addition of potassium, no other parameters other than the potassium density is changed while we are depositing potassium. As such, the measured change in resistivity corresponds to the resistivity added solely by potassium. After addition of potassium, the change in resistivity is calculated to be $\Delta \rho = 21.9 \pm 2.6$~k$\Omega$/$\mu$m with $2.2 \pm 0.1$ potassium atoms/$\mu$m for holes and$\Delta \rho = 6.5 \pm 2.3$~k$\Omega$/$\mu$m with $29.6 \pm 0.4$ potassium atoms/$\mu$m for electrons.

Figures 2c-d show the added resistivity as a function of gate voltages for holes and electrons at the same potassium densities as in Figures 2a and 2b. We find that determining resistivity is not possible for lower values of $V_g-V_{\rm onset}$ for both holes and electrons due to the appearance of a nonlinear dependence of the resistance on length. The nonlinearity is likely due to the contribution from the Schottky barrier at the contacts. The window for the linearity is significantly smaller for electrons. Within the windows where a linear dependence is observed, the behavior is consistent with the Coulomb scattering picture discussed above: the positively charged adsorbate creates a repulsive potential barrier for holes, and the sensitivity of the scattering strength to hole energy indicates that the imposed barrier height is comparable to the hole energy. For electrons, the potential barrier is attractive (i.e., it has the form of a well). The distance in energy between the bottom of the well and the electron energy is large and, as a result, for electrons, the magnitude of the scattering is small, with a very weak dependence on energy. We now discuss the resistivity added by potassium at high electron and hole energies, sufficient to minimize the effect from the Schottky barriers at the contacts.

\begin{figure}
\centering
\includegraphics[width=4.25cm]{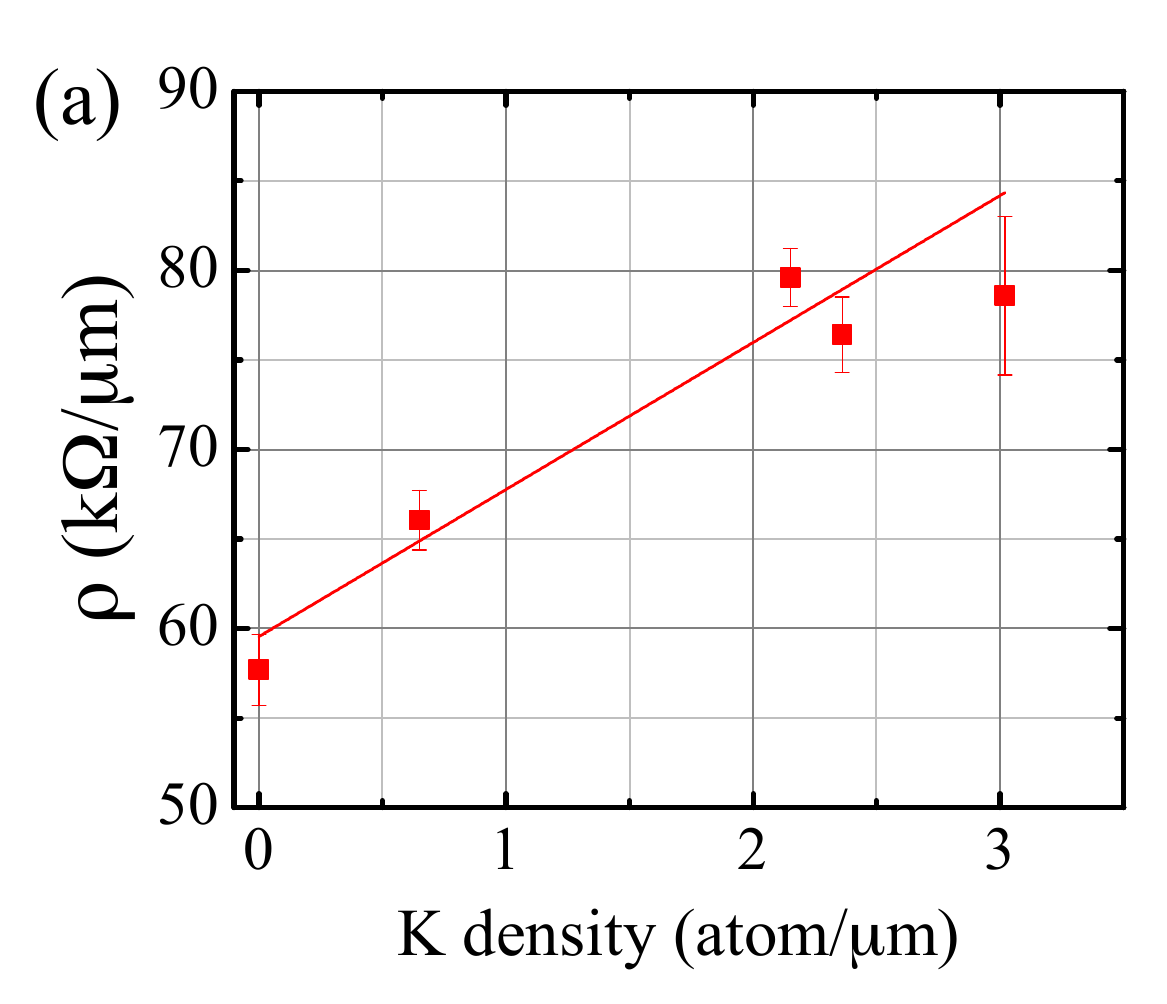}
\includegraphics[width=4.25cm]{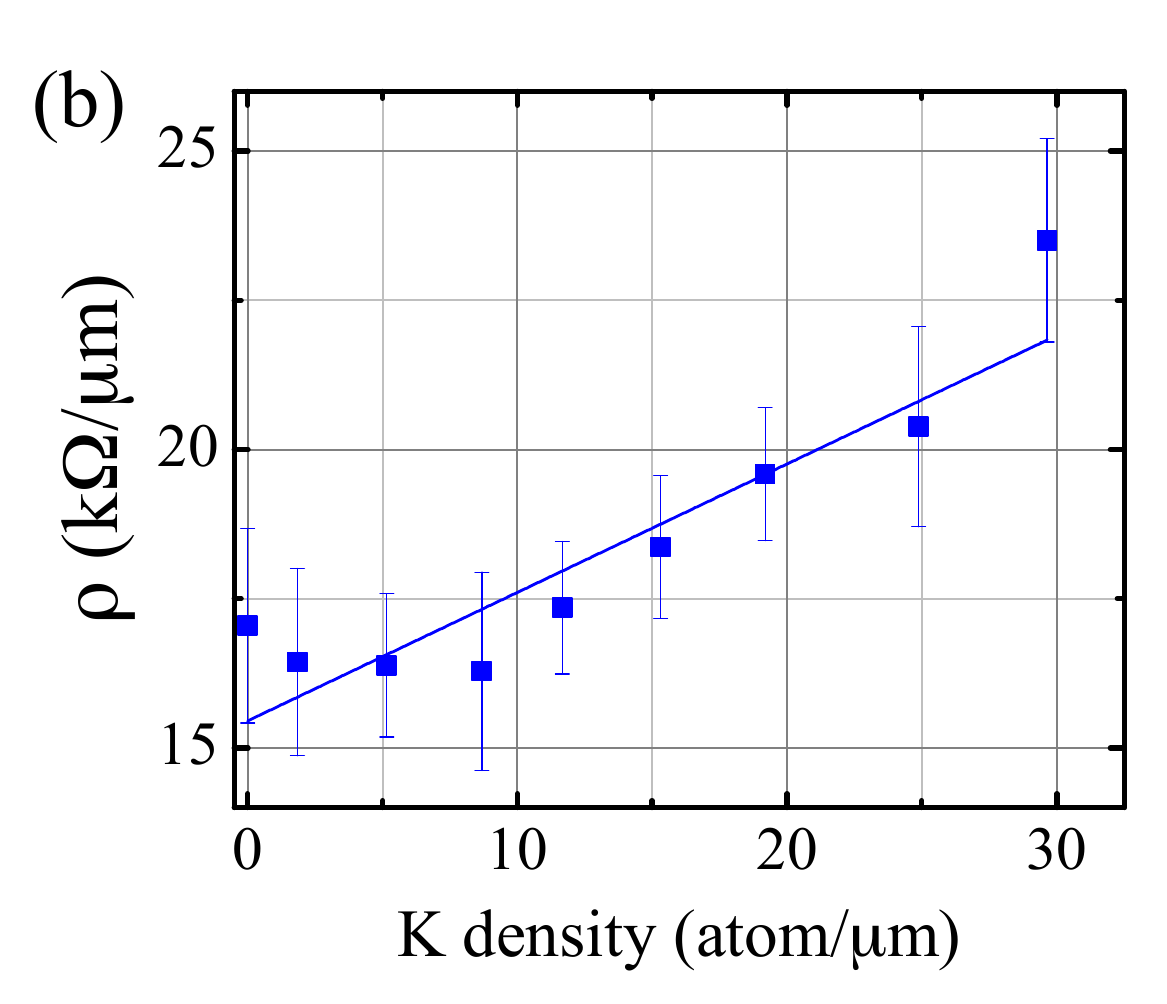}
\caption{(Color online) Resistivity as a function of potassium
  coverage for $V_g -V_{\rm onset} = -60$~V (hole) and $60$~V
  (electron).}
\label{fig:3}
\end{figure}

Figure 3 shows the measured resistivity at $V_g-V_{\rm onset}$ = $\pm 60$~V as a function of potassium density. At this energy, the contribution from the Schottky barrier from the contacts is minimal as evidenced by the linearity in resistance as a function of nanotube length. The added resistivity due to potassium remains close to linear with dosing within the error of measurements, indicating that potassium largely behaves as a diffusive and uncorrelated scatterer even at the maximum potassium densities of 3 atoms/$\mu$m for holes and 30 atoms/$\mu$m for electrons. Such adherence of the diffusive, semiclassical behavior is consistent with the phase coherence length of nanotubes being less than 100~nm for temperatures above 10~K \cite{ref35, ref36, ref37, ref38, ref39}, considerably smaller than the shortest segment measured. Furthermore, the linear dependence also indicates that potassium does not cluster. From the linear fit, the scattering strength of potassium is found to be $8.2 \pm 1.3$ k$\Omega$/atom for holes and $0.22 \pm 0.03$ k$\Omega$/atom for electrons. The scattering is 37 times larger for holes than for electrons. An asymmetry is expected because a potential barrier is a more effective scatterer than a potential well. The magnitude of the difference between electrons and holes depends on the scattering potential induced by the adsorbate. Thus, this measurement provides an essential experimental input to theory to determine the characteristics of this potential.

\begin{figure}
\centering
\includegraphics[width=4.25cm]{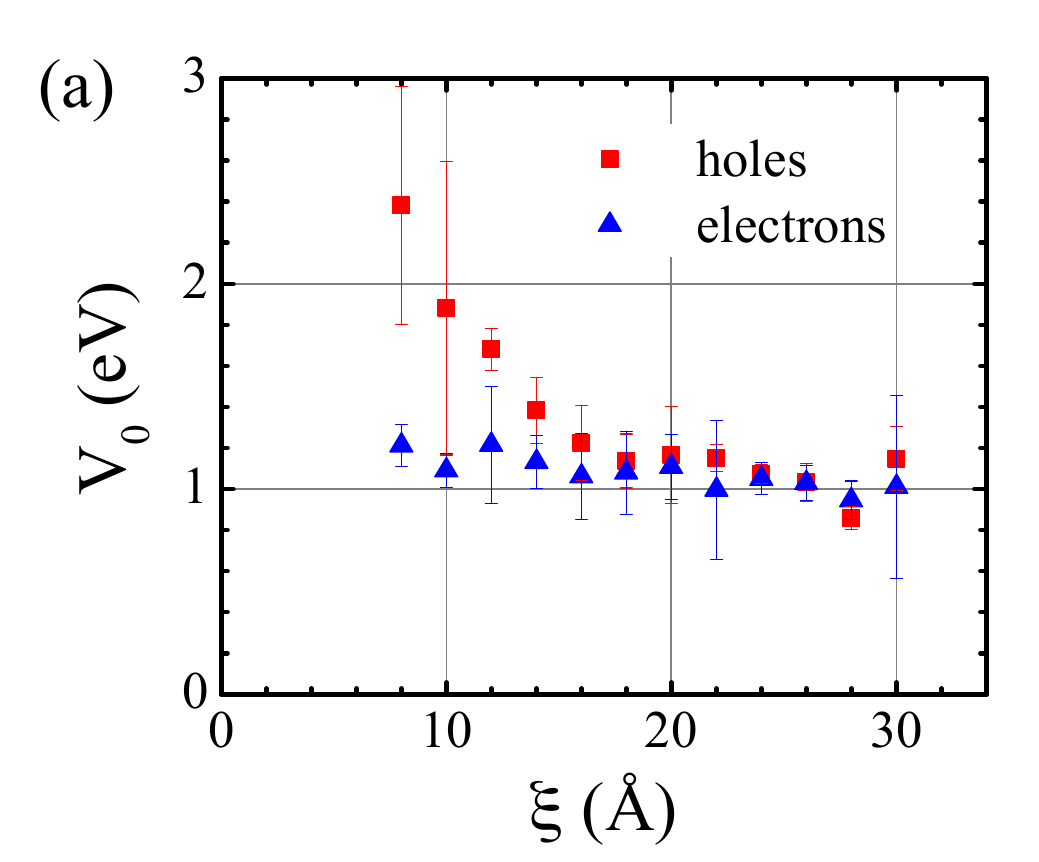}
\includegraphics[width=4.25cm]{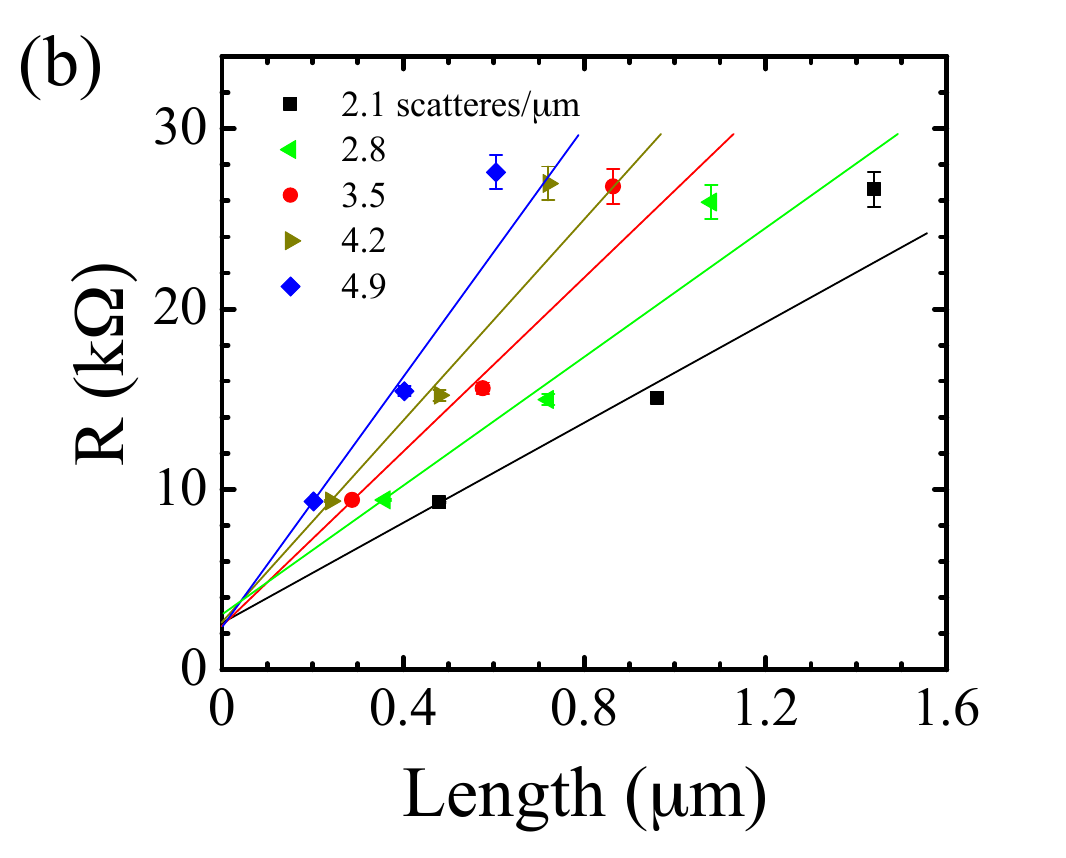}
\includegraphics[width=4.25cm]{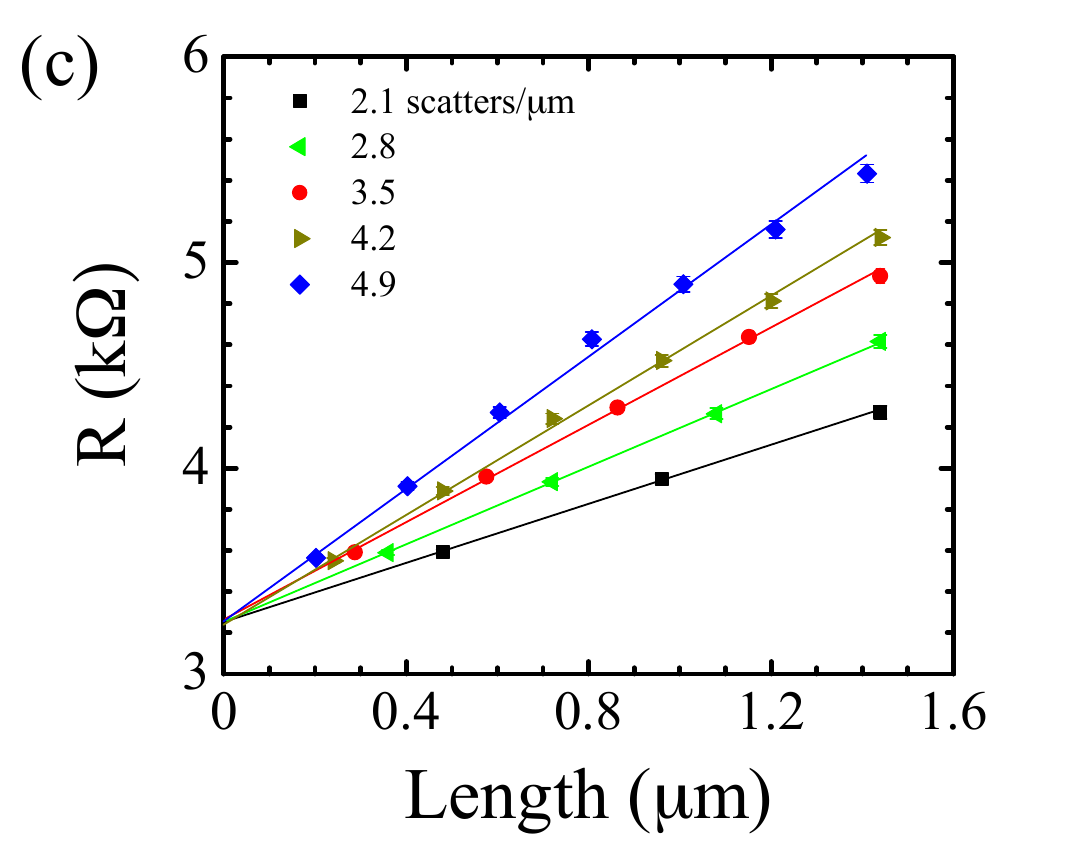}
\includegraphics[width=4.25cm]{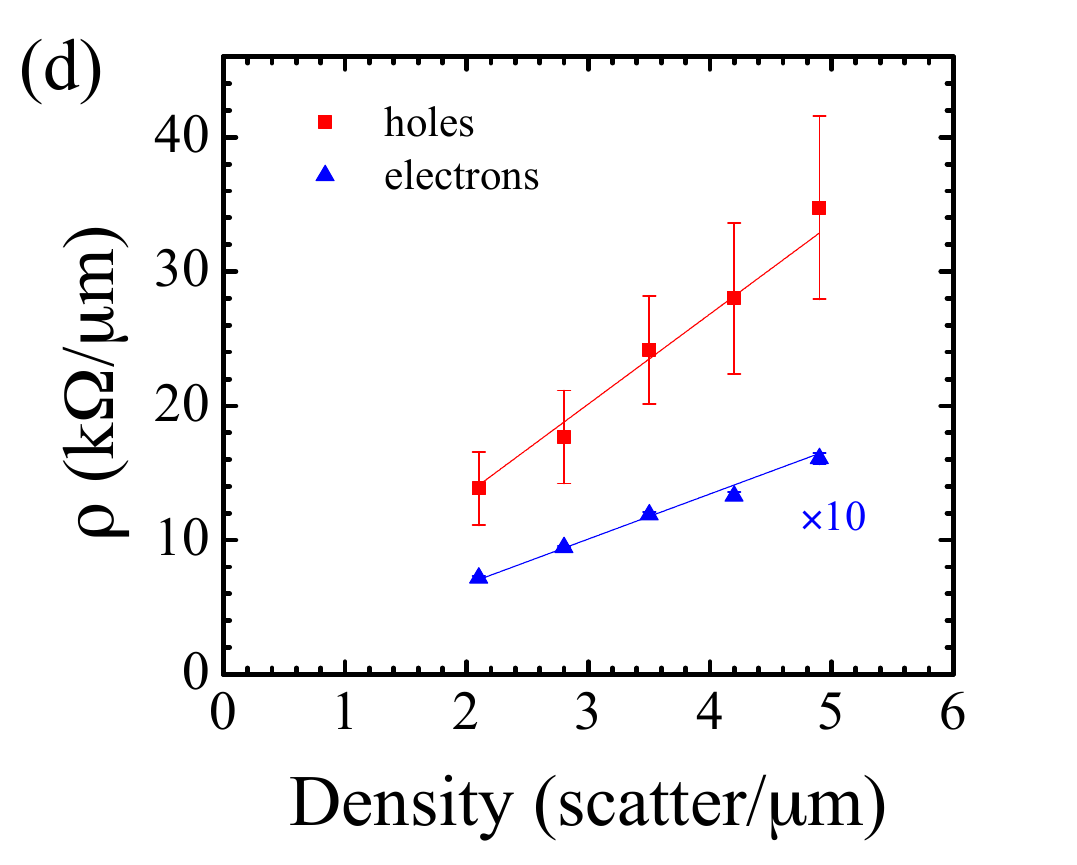}
\caption{(Color online) (a) Plot of the interrelation between $V_0$ and $\xi$ when the numerical value of the resistance per scatterer is fixed to the experimental value. Numerically evaluated average resistance versus length plot with different impurity density for (b) holes and (c) electrons. The energy is at $E = \pm 1.5$~eV with $V_0 = 1.1$~eV and $\xi = 20$ \AA . (d) Plot of the dependence of the resistivity on scatterer density. The results for electrons have been multiplied by a factor of 10 for a better clarity.}
\label{fig:4}
\end{figure}

We performed numerical transport calculations for a (7,6) nanotube using the recursive Green's function technique \cite{ref39} combined with the zero-temperature Landauer-B\"uttiker conductance formula. The strength of the scattering potential imposed by potassium was determined by a direct comparison of the numerical calculation with the experimental data. We started from a pristine single-band tight-binding model and added Gaussian potential scatterers of the form $V_i({\bf R}) = V_0 \exp\left(-|{\bf
  R} - {\bf R}_i|^2/\xi^2\right)$, uniformly distributed along the nanotube. Here ${\bf R}_i$ denotes the lattice location of the scatterer, $V_0$ its (positive) strength, and $\xi$ represents the scattering potential range. As expected, we found that the resistance of the nanotube varies considerably with $V_0$ and $\xi$. We adopted the following procedure in order to find values for these parameters. First, we evaluated the change in average resistance when a single scatterer was added to a short nanotube segment at random locations. Such simplified model enables rapid exploration of a wide range of values for $V_0$ and $\xi$. Second, for a fixed value of $\xi$, we varied $V_0$ until the change in average resistance at $E = \pm 1.5$~eV matched the corresponding experimental value within its numerical uncertainty. The result is shown in Figure 4a. The data points for electrons and holes differ substantially for short scatterings ranges, indicating inconsistency with the experimental data. The data points eventually begin to converge at increasing values of ξ before starting to separate again. As such, our analysis indicates that $\xi = 18 \sim 28$ \AA\ and $V_0 = 1.0 \sim 1.1$~eV are the choice of parameter values producing most consistent results with the experimental data at $E = \pm 1.5$~eV.

Using the ranges of values identified for $\xi$ and $V_0$ using the single-scatterer calculation, we performed more in-depth calculations that closely resemble the experiment. We evaluated the linear conductance for a wide range of nanotube lengths and scatterer concentrations, averaging each case over 600 random samples to wash away fluctuations due to phase-coherent interference. The scatterer concentration was varied within a range that kept transport diffusive (ohmic) and avoided Anderson localization of carriers. For this reason, the variation ranges for electrons and holes were different. The nanotube resistivity was obtained numerically following a procedure similar to that adopted in the experiments, namely, by varying the nanotube length (see Figures 4b and 4c for a typical determination of the resistivity). The scatterer resistance was then determined by considering the change of the average resistivity with scatterer density (Figure 4d). We find that the values of $\xi = 20$~\AA\ and $V_0 = 1.1$~eV for the spatial extent and the amplitude parameters of the impurity potential produce the closest results to the experimental values at the reference energies $E = \pm 1.5$~eV, yielding a scattering strength of $6.71 \pm 0.13$ k$\Omega$/scatterer for holes and $0.357 \pm 0.003$ k$\Omega$/scatterer for electrons, close to the experimentally observed values. A finer match might be possible employing numerical techniques that systematically avoid Anderson localization. The values of $\xi$ and $V_0$ identified by our theoretical analysis are significantly larger than those calculated for doped graphene where screening is expected to be stronger \cite{ref3}. Such weak screening even in the second subband, attested by the long scattering potential range, defies expectations from the previous calculations \cite{ref11, ref12, ref13} on the electron-electron screening. Therefore, our results suggest that existing theory on the electron-electron screening is inadequate for understanding screening of adsorbate-induced potentials in nanotubes.

In conclusion, we determined the resistance added by a potassium atom on a semiconducting single-walled nanotube of known chirality by measuring the resistivity added by adsorbates as a function of coverage in the diffusive transport regime. We found that the scattering strength of potassium is electron-hole asymmetric, with holes being more strongly scattered than electrons. The measurement of the scattering strength of an individual adsorbate on the nanotube allowed the determination of the depth and spatial extent of the scattering potential induced by a model Coulomb adsorbate. Our results represent a novel connection between experiment and theory on the study of adsorbate-induced scattering in nanotubes and pave the way for the fundamental science and the rational engineering of nanotube-based sensors.


{\em Acknowledgments}.--This work was supported by the National
Science Foundation under the Grants No. 1006230 and 1006533.



\end{document}